\documentclass[12pt, letterpaper]{article}

\usepackage{xspace}
\usepackage{amsmath} 
\usepackage{graphicx}
\usepackage[utf8]{inputenc}
\usepackage{booktabs}
\usepackage{hyperref}
\usepackage{aas_macros}

\usepackage[top=1in, bottom=1.5in, left=1in, right=1in]{geometry}

\graphicspath{{./figures/}}

\newcommand{\superk}  {Super\nobreakdash-K\xspace}

\title{LSST Target of Opportunity proposal for locating a core collapse
  supernova in our galaxy triggered by a neutrino supernova alert}
\author{ C.W. Walter, D. Scolnic, A. Slosar}
\date{ November 2018}

\begin{document}

\maketitle

\begin{abstract}

  A few times a century, a core collapse supernova (CCSN) occurs in
  our galaxy. When such galactic CCSNe happen, over 99\% of its
  gravitational binding energy is released in the form of neutrinos.
  Over a period of tens of seconds, a powerful neutrino flux is emitted
  from the collapsing star.  When the exploding shock wave finally
  reaches the surface of the star, optical photons escaping the
  expanding stellar envelope leave the star and eventually arrive at
  Earth as a visible brightening.

  By combining the multi-messenger signal from optical, neutrino, and
  gravitational waves, we afford an unprecedented opportunity to learn
  about the astrophysics of these rare objects. Carefully measuring
  the optical light curve of the explosion will give critical
  information about the size and composition of the progenitor star
  and help understand the dynamics of the explosion.

  Crucially, although the neutrino signal is prompt, the time to the
  shock wave breakout can be minutes to many hours later.  This means
  that the neutrino signal will serve as an alert, warning the
  optical astronomy community the light from the explosion is coming.  
  Quickly identifying the location of the supernova on the sky and
  disseminating it to the all available ground and spaced-based
  instruments will be critical to learn as much as possible about the
  event.

  Some neutrino experiments can report pointing information for these
  galactic CCSNe. In particular, the Super-Kamiokande experiment can
  point to a few degrees for CCSNe near the center of our galaxy.  A
  CCSN located 10~kpc from Earth is expected to result in a pointing
  resolution of the order of $3^\circ$.  LSST's field of view (FOV) is
  well matched to this initial search box.  LSSTs depth is also
  uniquely suited for identifying CCSNe even if they fail or are
  obscured by the dust of the galactic plane.

  This is a proposal to, upon receipt of such an alert, prioritize the
  use of LSST for a full day of observing to continuously monitor a
  pre-identified region of sky and, by using difference imaging,
  identify and announce the location of the supernova. In this
  proposal, we propose to use one night (approximately 0.03\% of the
  survey period) if a galactic supernova occurs.  Based on estimates
  of the rate of such CCSNe there is approximately a 20\% chance that a
  CCSN will occur during the survey period.
  
\end{abstract}

\newpage

\section{White Paper Information}

\noindent
Authors: \\

\noindent
Chris Walter - chris.walter@duke.edu \\
Dan Scolnic - daniel.scolnic@duke.edu \\
An\v{z}e Slosar - anze@bnl.gov \\

\noindent
Categorization: 
\begin{enumerate} 
\item {\bf Science Category:}  Exploring the transient optical sky
\item {\bf Survey Type Category:}  Target of Opportunity observation
\item {\bf Observing Strategy Category:}  A single night, continuous
  observation strategy focused on one few degree field pointing as given by
  a neutrino supernova alert trigger.
\end{enumerate}  

\clearpage

\section{Scientific Motivation}
\label{sec:motivation}

No visible galactic CCSNe have been seen and measured in the modern
scientific era. They are only thought to occur two or three times a
century~\cite{1994ApJS...92..487T, 2001ASSL..264..199C}.  The closest
modern CCSN we have observed was SN~1987A in the LMC.  Using an alarm
from neutrino detectors as a trigger, LSST can quickly identify and
characterize a galactic CCSN and then notify the rest of the
astronomical community which can then track it in multiple wavelengths
from the ground and space.  The early identification of the supernova
optical counterpart will be key in an extensive program of
multi-messenger astronomy.

How a neutrino-based supernova alarm can be used to prepare for the
arrival of an optical signal in LSST can be understood from the basic
sequence of CCSN formation.  As massive ($ > 8 M_\odot$ ) stars
approach the end of their lives, they begin to run out of the hydrogen
which has been fueling their fusion burning.  The resulting loss of
pressure results in a contraction of the star, raising its temperature
until the burning of the helium produced in the previous hydrogen
fusion can begin. This cycle repeats, next burning carbon, neon,
oxygen, and silicon until finally a core of iron remains. Upon
reaching the Chandrasekhar mass, this iron core collapses.  Not until
nuclear densities does the collapse halt, the supernova shock form,
and the neutrinos begin to stream towards Earth.  At this point the
ultimate fate of the star is determined by a battle between the
accreting matter and the intense outward pressures from heat,
neutrinos, and turbulence. If the former wins, a black hole is formed,
if the latter wins, the supernova shock drives out into the star
starting an explosion.  Depending on the size of the star, only
minutes to many hours later will the shock wave actually break out of
the stellar envelope and become visible as a supernova explosion to
optical telescopes.

Studying a galactic supernova in detail is an amazing opportunity to
do multi-messenger astronomy~\cite{2016MNRAS.461.3296N}.  Neutrino,
gravitational wave, and optical signals all tell us something unique
about the system.  The supernova converts the binding energy of the
star into energy and over 99\% of it is released in the form of
neutrinos.  Crucially, all of these neutrinos escape the star in the
first several tens of seconds of the explosion. Much can be learned
about the dynamics of the explosion by studying the neutrino signal.
Even the formation of a black hole, where the neutrino signal will be
abruptly terminated, should be visible~\cite{2011ApJ...730...70O,
  2017hsn..book.1555O}.

Additionally, studying the optical explosion signal in detail from the
beginning of the explosion will probe how the explosion proceeds and
will give crucial insights into the character, composition and size of
the progenitor star.  This will allow us to better understand the
final stages of stellar evolution and the environment that exists as
the collapse begins.  Examples of strategies to constrain the
progenitor characteristics and explosion dynamics from light curves
include~\cite{2010ApJ...725..904N, 2017NatPh..13..510Y,
  2018ApJ...856..146A}.  Some models of black hole formation even
include a much reduced, but possibly visible, electromagnetic
signal~\cite{2013ApJ...769..109L}. With the progenitor information
gained from the light curve we can connect what is happening on the
outside of the star to what is happening on the inside as measured by
the neutrino and gravitational wave signal.
 
A world-wide network of neutrino detectors including Super-Kamiokande
(\superk)~\cite{2003NIMPA.501..418F} have prompt alarms to alert the
world if a supernova neutrino burst has been seen. All of these
experiments are also networked together into a system known as SNEWS
(The Supernova Early Warning System; website and more information at
\url{https://snews.bnl.gov})~\cite{2004NJPh....6..114A}. SNEWS does a
blinded coincidence between experimental alerts sending out an
automated announcement to the GCN if more than one neutrino experiment
has seen a burst of neutrinos, and is also capable of sending a
individual confirmed alerts from a single experiment.  Currently,
\superk is the only running experiment with pointing ability, and we
concentrate on its performance for this proposal.  The neutrino
interactions inside of \superk from a CCSN consist of inverse beta
decay (IBD) interactions on nuclei with only weak directional memory
and also scattering on atomic electrons which point to their source
quite well.

For a more detailed description of the expected fluxes from each
neutrino type and neutrino interactions expected in detector refer to
the review~\cite{2012ARNPS..62...81S} and
section~\ref{sec:interactions} below.
Figure~\ref{fig:SK-realtime-monitor-pointing} taken
from~\cite{2016APh....81...39A} shows a simulated example of
interactions from a supernova near the galactic center with its
direction reconstructed.  In the figure, the electron scattering
interactions are in red, the IBD interactions in blue.  The fluxes
(and resulting pointings) are model dependent, but studies have shown
that for a CCSN located 10~kpc away it is possible to determine the
direction of the star to within about 3-5
degrees~\cite{2016APh....81...39A}.  Closer or more luminous CCSNe
will have better pointing.  The expected time delay ranges from
minutes to more than a day depending on the mass of the progenitor
star. Figure~\ref{fig:delay-times} taken
from~\cite{2013ApJ...778...81K} shows the range of expected times.
The SN~1987A progenitor was thought to be a blue giant with a time
delay of around 3 hours.

LSST is particularly well suited to do the initial CCSN
identification.  LSST's large 3.6 degree FOV is well matched to the
initial search region that would come from \superk.
LSST can continually collect exposures in the region until the
explosion is seen. Next is LSST's depth.  There are other wide-field
surveys that might easily see the supernova if it is bright.  But,
even with a large neutrino signal, the optical signal from the CCSN
could be quite dim.  This is a place where LSST will make a
particularly vital contribution.  Recent work has estimated that a
supernova located in the disk obscured by dust could be as dim as
magnitude 25~\cite{2016MNRAS.461.3296N}.  The explosion could also
fail or form a black hole~\cite{2011ApJ...730...70O,
  2017hsn..book.1555O}.  The expected range of brightnesses are
explored in~\cite{2016MNRAS.461.3296N},
figure~\ref{fig:multimessenger-comparison} taken from that paper shows
the reach of LSST for the dimmest supernova compared to other
facilities.

Finally, to quickly identify the CCSN, deep image templates of the
area will be necessary for image subtraction and candidate
identification.  Likely the depth of these templates will be the main
limiting factor for identifying faint candidates.  LSST should have
templates across the sky after Y1.  The depth $\times$ area on average
across the sky will be deeper for LSST than any other survey.

Identifying and studying a galactic supernova would be a scientific
gold-mine for astronomy and particle physics.  The merit of enabling
these studies is very high. The impact on running is minimal.  Using
the rate of three per century there is a 20\% chance that we would
receive a neutrino alarm. In that case we advocate for a strategy of a
full day of observing with follow-up over next few days to ensure the
candidate.  Depending on how bright it is, we would quickly hand off
to other ground and space-based telescopes.  It would take
approximately 0.03\% of the survey's time to contribute to a major
discovery.

\newpage

\begin{figure}
  \begin{center}
    \includegraphics[width=3.9in]{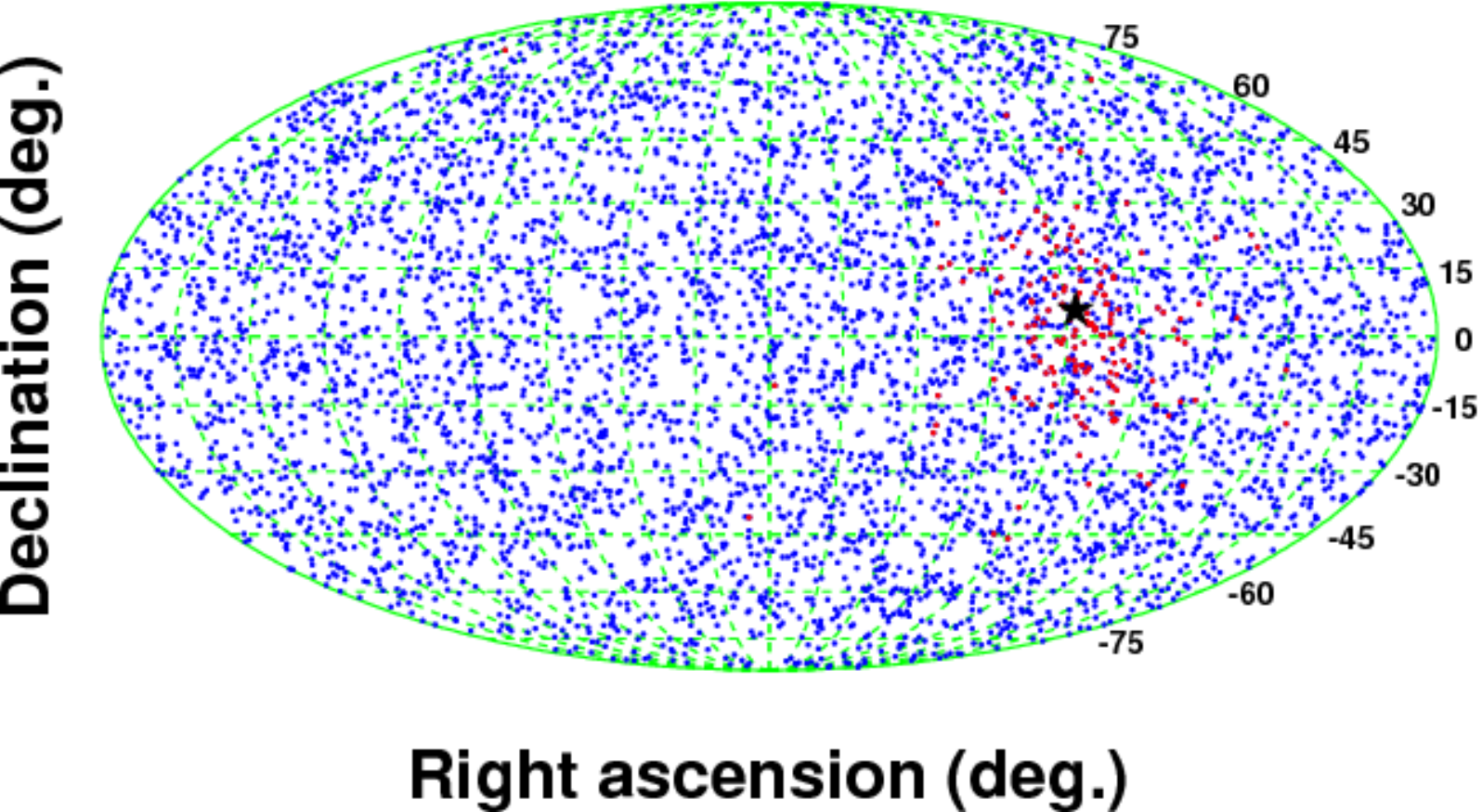}
    \caption{Figure 7 from~\cite{2016APh....81...39A} shows how the
      CCSN location is determined. A simulated set of neutrino
      interactions from a supernova 10~kpc from Earth are displayed.
      The blue points are the IBD events, the red the electron
      scattering events.  The reconstructed direction after fitting
      the peak is shown with a star. See
      reference~\cite{2016APh....81...39A} for more detailed
      information. }
    \label{fig:SK-realtime-monitor-pointing}
  \end{center}
\end{figure}

\begin{figure}
  \begin{center}
    \begin{minipage}[b]{3.1in}
      \includegraphics[width=3.0in]{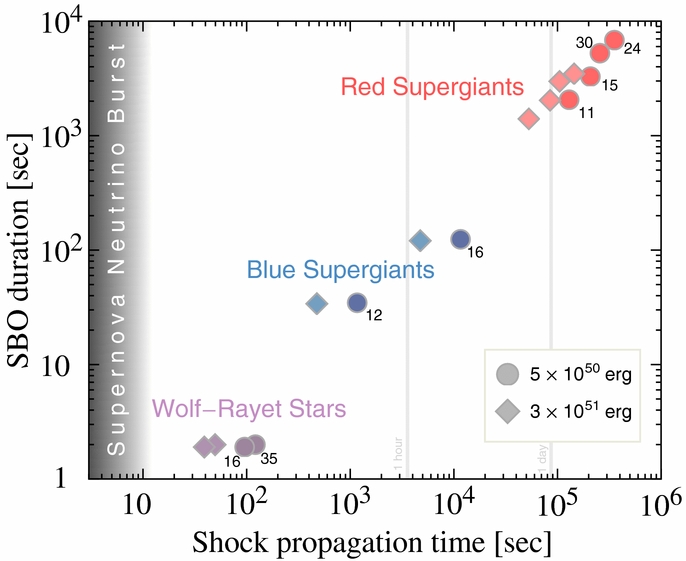}
      \caption{Figure 2 from~\cite{2013ApJ...778...81K} shows
        calculated shock propagation times for different classes of
        stellar progenitors.  The propagation time sets the delay
        between the prompt neutrino signal and the optical signal
        which appears when the shock breaks out of the stellar
        envelope.  See reference~\cite{2013ApJ...778...81K} for more
        details. \textcopyright~AAS. Reproduced with permission.}
      \label{fig:delay-times}
    \end{minipage}
    \hspace{0.2in}
    \begin{minipage}[b]{3.1in}
      \includegraphics[width=3.0in]{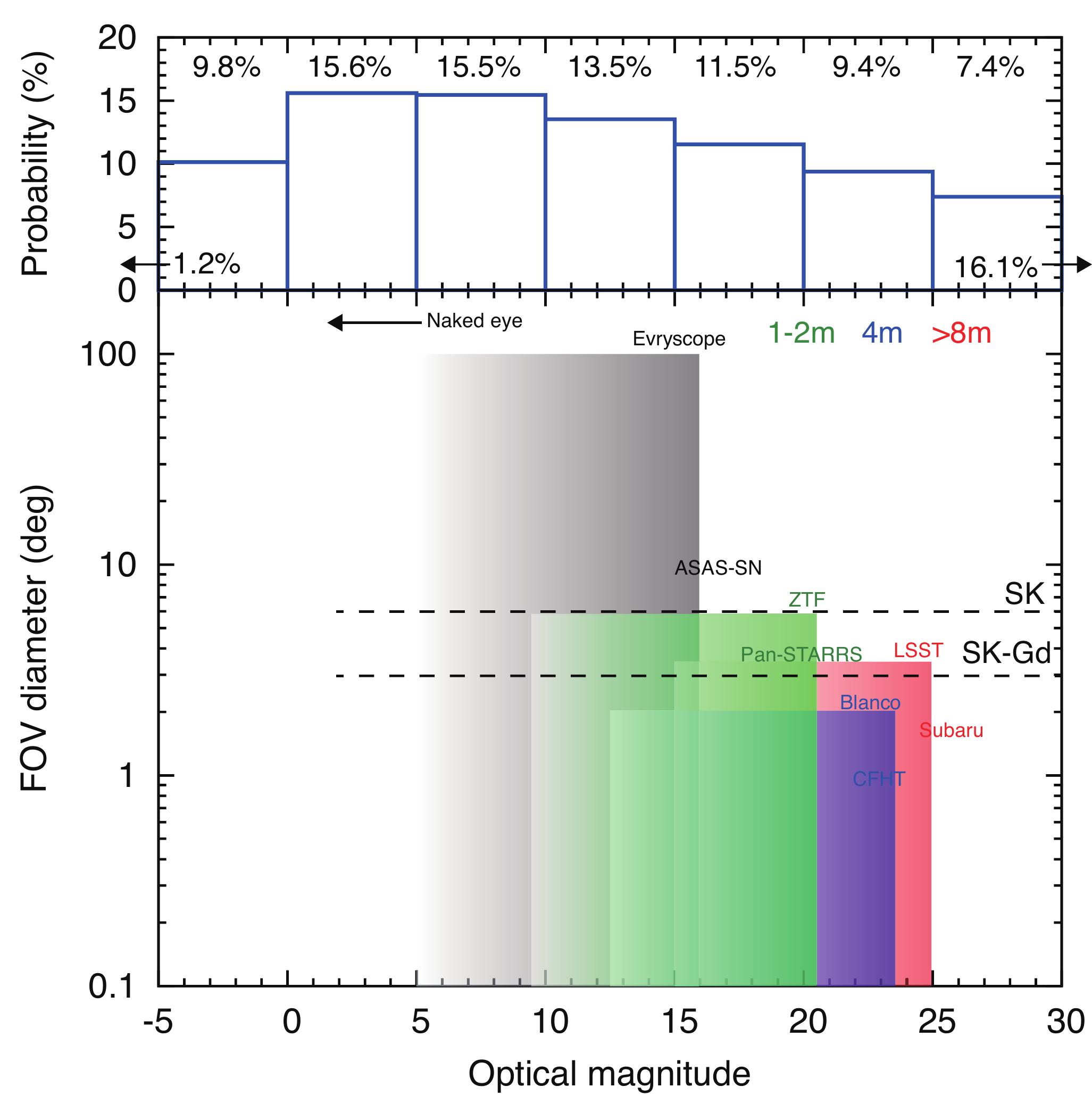}
      \caption{The top panel of figure 9
        from~\cite{2016MNRAS.461.3296N} shows the predicted
        dust-attenuated peak magnitudes and corresponding percentage
        fraction of all CCSNe.  The bottom panel shows typical
        magnitudes and FOV for various optical telescopes including
        LSST.  The SK expected resolution is also indicated.  See
        reference~\cite{2016MNRAS.461.3296N} for more details.}
      \label{fig:multimessenger-comparison}
    \end{minipage}
  \end{center}
\end{figure}

\clearpage

\newpage
\section{Technical Description}

In order to describe the survey strategy for footprint, tiling method,
depth, and observation frequency it is important to understand the
expected time delay and pointing signal along with the form of the alarm
signal.  Those are first summarized here.

\subsection{Expected Time Delay}

The time delay between the neutrino alarm and the light signal
reaching LSST can range from minutes in the case of Wolf-Rayet stars,
to hours for blue supergiants, all the way up to a day or two for the
largest red supergiants.  The time is set by the radius of the star
when the collapse happens, as that sets the distance that the shock
wave must travel.  Figure~\ref{fig:delay-times} taken
from~\cite{2013ApJ...778...81K} shows how the breakout time of the
shockwave varies for different classes of stellar objects. Table~2
in~\cite{2015ApJ...814...63M} calculates the delay times for various
modeled red supergiants.  Fractions of CCSN classification taken
from~\cite{2017PASP..129e4201S} suggest that Wolf-Rayet stars will have
a delay of 1 to 10 minutes and are approximately 30\% of CCSN, blue
supergiants a few hours and 15\%, and red supergiants one to two days 
and 55\%.

\subsection{Neutrino Interactions}
\label{sec:interactions}

The expected pointing resolution in a water Cherenkov detector will
scale with the number of interactions detected.  As explained in
section~\ref{sec:motivation} a set of electron scattering interactions
which point back to the supernova will be sitting on top of a
background of poorly-pointing interactions from the IBD neutrino
captures. For a supernova located 10~kpc from the Earth (the galactic
center is approximately 8~kpc away) the order of 10,000 neutrino
interactions are expected in \superk.  Supernova that are closer or
further away will have their fluxes scaled simply by scaling to their
distance with a factor of $1/r^2$.  Although something like 10,000
interactions are typical, expected fluxes are found to have a range of
values by different simulation groups.

Some detectors can only report the time and size of a neutrino burst,
and do not have the ability to supply directional information.  In the
near future, there will be new detectors with pointing ability and the
SNEWS system will also add pointing information utilizing
intra-detector timing by using the fact that travel times across the
Earth from multiple detectors (such as \superk and IceCube) can
triangulate the CCSN position~\cite{2018JCAP...04..025B} and
[N. Linzer and K. Scholberg in preparation].  However, currently,
\superk is the only running experiment with pointing ability, and we
use its performance for this proposal.

Most of the neutrino interactions in the water of the \superk
experiment are inverse beta decay (IBD)
$$ \overline{\nu}_{e}+ p \rightarrow e^{+} + n , $$
where a neutrino is
captured by a proton resulting in a positron (which is detected
through its Cherenkov radiation) and a neutron.  The positron in this
reaction carries only very weak directional memory of the incoming
neutrino. However, a few percent of the neutrino interactions proceed
through atomic electron scattering:
$$\nu + e^{-} \rightarrow \nu + e^{-} .$$
Unlike the IBD reactions,
these atomic electron scattering interactions {\bf do} point back to
the supernova well.  In order to find the direction of the CCSN a fit
is done to the two components of the signal resulting in an inferred
position and error region on the sky for the supernova position.

\subsection{Expected pointing resolution}

Given a number of interactions, pointing to the supernova using the
electron scattering signal only given is by the kinematics of the
interaction and is found to give a resolution of roughly
$$ \Delta \theta = \frac{30^\circ}{\sqrt{N}.}, $$
where N is the number of electron-neutrino scattering interactions and
the angular resolution is a half-opening
angle~\cite{2012ARNPS..62...81S}.  With a few percent of 10,000
interactions from a 10~kpc supernova being electron scattering events,
this tells us that we should expect a rough pointing resolution of
close to $1.5^\circ$.   However, this resolution will be degraded
due to the fact that the scattering signal is on top of the IBD
background. Closer supernovae will have higher numbers of
interactions with better pointing, and those further away will have
their resolution decreased.

\begin{figure}
  \begin{center}
    \includegraphics[width=4.0in]{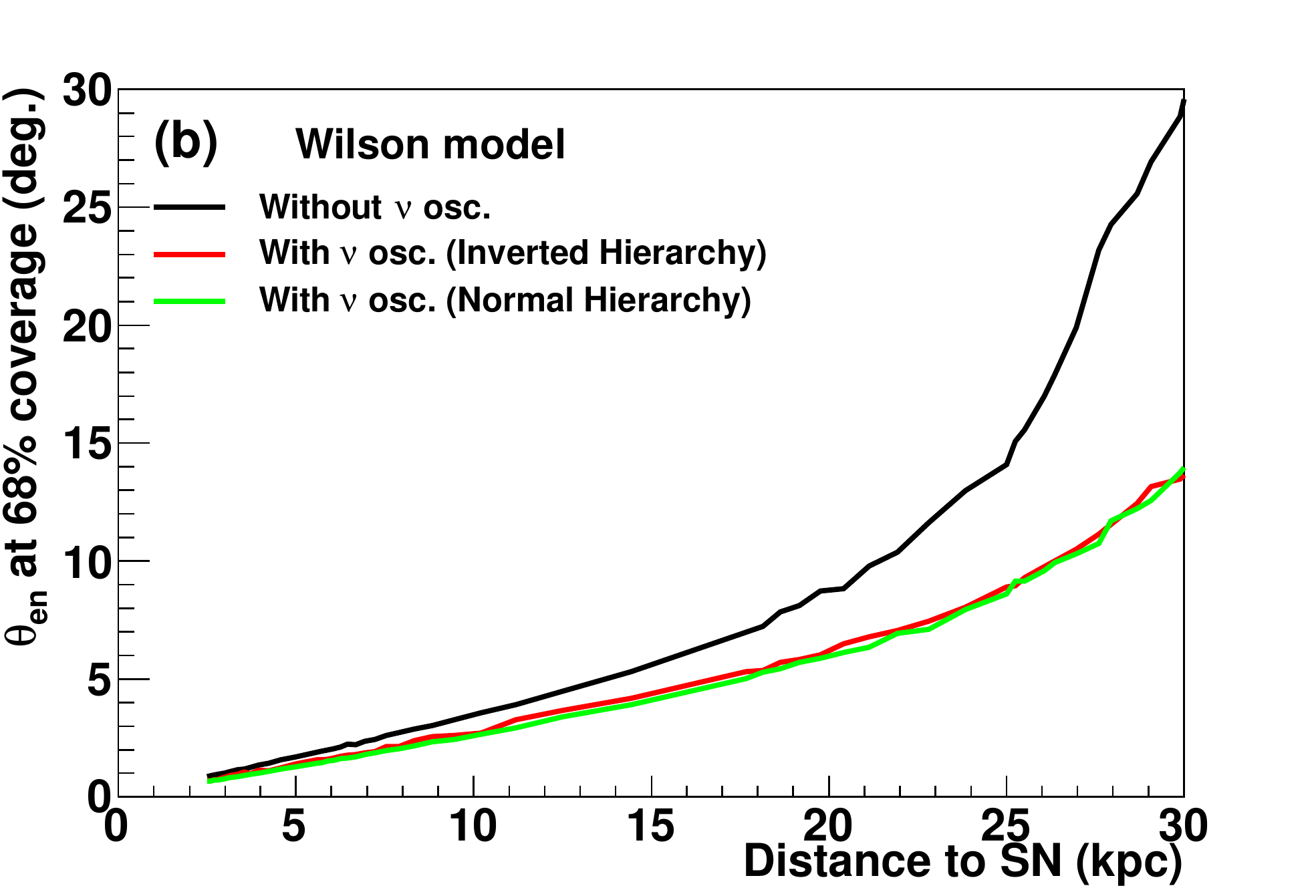}
    \caption{Figure 9b from~\cite{2016APh....81...39A} shows the 68\%
      $\Delta \theta$ reconstructed by the Super-K algorithm as a
      function of distance to the CCSN from the Earth in kpc using
      simulated data in Super-K from a particular (the Wilson) model.
      With neutrino oscillations properly accounted for, a supernova
      from near the Galactic center results in an opening angle
      resolution of about $3^\circ$.}
    \label{fig:SK-realtime-pointing-resolution}
  \end{center}
\end{figure}

A careful study by the \superk collaboration
in~\cite{2016APh....81...39A} plots a 68\% opening angle coverage as a
function of distance for a few flux models with a realistic fitting
procedure to accommodate the IBD
background. Figure~\ref{fig:SK-realtime-pointing-resolution} shows how
the pointing is expected to scale as a function of distance with
neutrino oscillations taken into account for one of the models. At
10~kpc the pointing is near $3^\circ$.

By the time the LSST survey begins we expect \superk to be doped with
0.02\% $\textrm Gd_2(SO_4)_3$~\cite{2004PhRvL..93q1101B}.  The
addition of gadolinium (which has a high neutron cross-section) to the
water will allow the neutron to be tagged in IBD events thus removing
a portion of the non-pointing background and improving the pointing
resolution to be closer to the formula above.

\subsection{Alert input}

In order to point LSST, the observatory control system (OCS) must
first receive information from the neutrino experiments that a light
from a galactic supernova is about to arrive.  Time is of the essence
since depending on the size and type of the star the breakout time
could range from minutes to many hours~\cite{2013ApJ...778...81K}.
There is currently more than one way to receive an alert. If this
proposal is selected LSST must work with the neutrino community to
ensure that the information that LSST needs is being promptly
transferred.

There are two broad classes of alerts to consider.  Each experiment
has the option to send its own alert to the astronomical community.
For example, if \superk determines a CCSN in our galaxy has occurred
it might send the following template-like text example to the Astronomers
Telegram:

\begin{verbatim}
Super-Kamiokande, a 50000 ton water Cherenkov imaging detector
situated 1000 meters underground in the Kamioka mine, Gifu, Japan, has
observed a neutrino burst from a nearby supernova.  Within a fiducial
volume of 22500 tons, preliminary results indicate 5227
neutrino-produced events have been detected with energies greater than
7.0 MeV. An SN1987A-like explosion would be expected to produce such a
signal in Super-Kamiokande if the progenitor star was located at a
distance between 7.55 and 10.36~kpc from Earth. These events were
observed over an interval of 17.9 seconds, with the first event
arriving at 2017 Nov 2.318437 UT. The estimated supernova direction is
R.A. = 110 (degrees) and Dec.= 6 (degrees), within 3.29, 4.72 and 5.62
degrees for, respectively, 68, 90 and 95% C.L. error circles. 
The probability to have the SN located within 2, 5, and 10 degrees 
of the central position is 0.36, 0.92 and 1.00, respectively.
\end{verbatim}

C.W. Walter is a member of both \superk and the LSST project and
\superk has expressed interest (in personal communications) to CWW to
supplying direct information to the LSST OCS in whatever form is most
appropriate if necessary.

For many years the neutrino community as a whole has being preparing
for a galactic supernova through the creation of the Supernova Early
Warning System (SNEWS).  SNEWS acts as a
broker and a blinded system to look for coincidences in time between
supernova alarms coming from different neutrino experiments. If one is
seen, then they alert the entire astronomical community through several
channels. This reduces the false coincidence rate to less than one
alert per century. SNEWS can also act as a broker passing alarms from
individual experiments to their alert system

SNEWS has several ways of making announcements to the community.  They
also give a direct connection to the IceCube experiment which benefits
from an external trigger.  A direct connection to LSST could also be
requested.  Current  alerts include announcements to the GCN (a
template is seen below):

\begin{verbatim}
TITLE:         GCN/SNEWS EVENT NOTICE
NOTICE_DATE:   Tue 26 Jun 18 16:00:08 UT
NOTICE_TYPE:   TEST COINCIDENCE
TRIGGER_NUM:   1000182
EVENT_RA:      Undefined (J2000),
              Undefined (current),
              Undefined (1950)
EVENT_DEC:     Undefined (J2000),
              Undefined (current),
              Undefined (1950)
EVENT_ERROR:   360.0 [deg radius, statistical plus systematic], 68.00% containment
EVENT_FLUENCE: 0 [neutrinos]
EVENT_TIME:    57601.00 SOD {16:00:01.00} UT
EVENT_DATE:    18295 TJD;   177 DOY;   18/06/26
EVENT_DUR:     0.00 [sec]
EXPT:          Detector_A Good, Detector_B Good, Detector_D Possible, Detector_E Good, Detector_F Possible, 
SUN_POSTN:      95.45d {+06h 21m 49s}  +23.34d {+23d 20' 30"}
SUN_DIST:      Undefined [deg]
MOON_POSTN:    257.26d {+17h 09m 02s}  -19.06d {-19d 03' 38"}
MOON_DIST:     Undefined [deg]
MOON_ILLUM:    98 [%]
GAL_COORDS:    Undefined,Undefined [deg] galactic lon,lat of the event
ECL_COORDS:    Undefined,Undefined [deg] ecliptic lon,lat of the event
COMMENTS:      SNEWS Event without RA,Dec coordinates.  
COMMENTS:      This is a Test COINCIDENCE notice.  It is NOT a Real event.  
COMMENTS:      This is a Test COINCIDENCE notice.  The EXPT labels have been anonymized.  
COMMENTS:         
COMMENTS:      RA,Dec fields undefined.  
COMMENTS:      For more information see:  
COMMENTS:
\end{verbatim}


\noindent
and email alerts such as the template below.

\begin{verbatim}
-----BEGIN PGP SIGNED MESSAGE-----
Hash: SHA1

- ---------------------------------------------
*** SNEWS ALERT ***
Coincidence rating: GOLD
Alarms in the coincidence:
Experiment: 5 LVD
Level: GOOD
Time: Jan 02 2006 22:34:37.000000000
Duration:   10.00
No. of signal events:    0.00
Right Ascension:    0.00
Declination:    0.00
Error:  360.10
- ---------------------------------------------
Experiment: 3 SNO
Level: GOOD 
Time: Jan 02 2006 22:34:37.000000000
Duration:   10.00
No. of signal events:    0.00
Right Ascension:    0.00
Declination:    0.00
Error:  360.10
- ---------------------------------------------
Experiment: 1 Super-K
Level: POSSIBLE 
Time: Jan 02 2006 22:34:37.000000000
Duration:   10.00
No. of signal events:    0.00
Right Ascension:    13.00
Declination:    -3.00
Error:  4.0
- ---------------------------------------------

For information, see web page http://snews.bnl.gov/
-----BEGIN PGP SIGNATURE-----
Version: GnuPG v1.4.9 (GNU/Linux)

iD8DBQFMhguY4A2qNGjfk/cRAp+DAKD2cFdN4aHZomU87XhhA2r7GalWcACgt/oM
ffObwWjd44FA6kx5gx/RLDQ=
=DtVE
-----END PGP SIGNATURE-----
\end{verbatim}

Currently, a fast alarm goes directly from \superk to the SNEWS system
with no human intervention.  However, that alarm does not contain
pointing information.  Now, that information is released only after a
virtual meeting of \superk collaborators to confirm the alarm.
However, it is recognized that this step slows down dissemination, so
discussion is starting on the best way to pass this information either
directly to projects like LSST or through systems like SNEWS.

LSST could decide to require multiple or single experimental alerts,
while taking the pointing system from \superk.  In the future, SNEWS
is also expected to supply pointing based on triangulation using
timing and this information could be combined with the electron
scattering signal~\cite{2018JCAP...04..025B} and [N. Linzer and
K. Scholberg in preparation].

\subsection{High-level description}

We propose a new LSST ``Supernova watch mode'' that would be triggered by a
neutrino supernova alert and would stay engaged for the rest of the
observing day.  Schematically the sequence of events would be as follows:

\begin{itemize}
\item Receive neutrino alert
\item Alert is immediately passed to relevant personnel on duty
\item If the telescope is not observing (because it is day time or bad
  weather), if the field is not currently visible, or if the candidate position is too near the Sun at all times, the relevant
  personnel can override the default strategy
\item If the field is visible, or when it becomes visible, preempt all
  the other observing plans
\item Note that if field rises hours after the first alert it might be possible that:
  \begin{itemize}
  \item The object has already been identified with high certainty in
    which case the operator on duty may decide for forfeit the
    opportunity
  \item The SN has already brightened, in which case it will be
    discovered in the first image, assuming good templates exist.
\end{itemize}

\item Verify up-to-date deep templates have been built and are accessible for that sky area in all filters.  If there are no recent deep templates available, the continuous exposures taken during the period wanting for the optical signal should be used to build one.  In this case, the sensitivity will increase the longer the time between the neutrino and optical signal.

\item Choose exposure, filter and dither plan:
\begin{itemize}
\item Filter and exposure plans should be pre-determined in
  conjunction with relevant experts. We do not currently possess a
  detailed recommendation. General guidelines are given below.
\item If the search region exceed the LSST pointing size, a dithering
  strategy covering 99\% of the area should be used.
\end{itemize}

\item Observe repeatedly in the region while continuously performing DIA analysis with the template for each short group of exposures and for the stacked nightly visit total at that time

\item If CCSN candidates are identified, pass this information to
  community and event brokers
\item Follow the light curve for the rest of the night of
  observation. This serves several purposes:
  \begin{itemize}
  \item It will help identify further candidates in case the first
    candidate was a mis-identification
  \item It puts the full light-curve on the same calibration basis,
    facilitating cross-calibration with other instruments which will
    presumably be following up.
  \end{itemize}
\item If no candidate is seen on the first night of observation at the
  LSST site, the CCSN may have failed, been obscured, or the
  progenitor star may be so large that the shock wave has not yet
  broken out of the stellar envelope.  A completely automatic response
  the first night is necessary, but we advise a protocol including
  consultation with experts and the rest of the community for the next
  night's observing plan.
\end{itemize}

\subsection{Footprint -- pointings, regions and/or constraints}

LSST can expect to receive an indicated region of the sky with a
target RA and Dec and (for example) and a half angle opening region
with a 68\% coverage.  We advocate covering 99\% of the localization
area with a strategy of dithered sampling that emphasizes regions with
higher probability based on the neutrino data prior.  The distance and
neutrino luminosity of the CCSN could result in pointing resolution
ranging from 1 to several degrees on the sky with a typical 10~kpc
supernova being localized to about a 6 degree FOV (twice the opening
angle resolution).

\subsection{Image quality}

The image quality is not relevant.  Exposures should be taken if at all possible.

\subsection{Individual image depth and/or sky brightness}

Single visits have a depth of 23.14,  24.54, 24.20, 23.65, 22.77,
21.92 in ugrizy.  Whether single exposures and differences are
effective in seeing the transient will depend on the optical brightness.
Trade offs in allocation of filter time are likely galactic coordinate dependent as they depend on dust, and an optimal strategy will require further study.

\subsection{Co-added image depth and/or total number of visits}

As this proposal covers a single night, final multi-year co-added depths are not relevant.

\subsection{Number of visits within a night}

This mode would be envisioned to completely take over the facility for
one dark period after the alarm arrives.

\subsection{Distribution of visits over time}

This proposal recommends and is relevant to a continuous exposure
strategy from as soon as the neutrino alert is received, through
identification and announcement and for the rest of the observing
night.  Possible later followup visits would depend on the nature of
the observed CCSN and the perceived need for LSST followup.

\subsection{Filter choice}

As the CCSN may be located in a high dust region, we prefer that the filter wheel contain the griz and y filters, as the y filter may be particularly useful in that case.

\subsection{Exposure constraints}

In Figure 7 of~\cite{2016MNRAS.461.3296N} Nakumura and all show that
the extincted optical signal in the visible for a galactic center CCSN
could range from magnitude 5 to 26. The apparent magnitude of the
light curve is a strong function of galactic position (shown in Figure
8 of the same paper).  A dynamic exposure scheme where the exposure
time is shortened for positions of little extinction, or
identification of a very bright object could be considered.

\subsection{Other constraints}

No other relevant constraints.  Clearly the target CCSN would need to be
visible to LSST to enter a supernova watch mode.  Depending on the
time of day, and time delay we might enter watch mode as soon as it
become dark.

\subsection{Estimated time requirement}

We expect no more than one night in the 10 year survey would be
affected by this program. With an expected rate of 3 per century there
is a 22\% chance one CCSN would be seen and a 3\% chance there could
be two.

\subsection{Technical trades}

Not relevant for this proposal.

\section{Performance Evaluation}

Not relevant for this proposal.  Performance would be measured in successfully
identifying and quickly notifying the community of the location of the
of the supernova.

\section{Special Data Processing}

A version of the DIA pipeline would need to be utilized to make the
initial identification.  As one of the advantages of leveraging LSST
for this work is its ability to see CCSNe which have been obscured by
dust in the galactic plane, we also advocate building a set of
templates in the galactic plane to use if the neutrino signal points
us to that region of the sky.

\section{Acknowledgments}

The authors would like to thank Kate Scholberg of Duke University for
information on the SNEWS system and general information on supernova
neutrino detection, Evan O'Connor of Stockholm University for
very useful information and references related to the modeling of CCSNe
and the properties of the progenitor stars and light curves, and the
Super-Kamiokande Spokesperson Masayuki Nakahata for information on
Super-K performance and plans.

\section{References}

\bibliographystyle{hunsrt} 
\bibliography{references}

\end{document}